# SIMPLE (SISYPHUS) DYNAMICAL MODEL OF THE COSMOLOGICAL CONSTANT

Vladan Panković

Department of Physics, Faculty of Sciences, 21000 Novi Sad,
Trg Dositeja Obradovića 4., Serbia

## Abstract

In this work we suggest (in a formal analogy with Linde chaotic inflation scenario) simple dynamical model of the dark energy or cosmological constant. Concretely, we suggest a Lagrangian dependent of Universe scale factor and scalar field (with constant and positive total energy density as cosmological constant). Then, Euler-Lagrange equation for Universe scale factor is equivalent to the second Friedman equation for the flat empty space with cosmological constant (in this sense our model is full agreement with recent astronomical observations). Also there is Euler-Lagrange equation for scalar field that includes additional "friction" term and negative first partial derivative of unknown potential energy density (this equation is, in some way, similar to Klein-Gordon equation modified for cosmic expansion in Linde chaotic inflation scenario). Finally, total time derivative of the (constant) scalar field total energy density must be zero. It implies third dynamical equation which is equivalent to usual Euler-Lagrange equation with positive partial derivative of unknown potential energy density (this equation is formally exactly equivalent to corresponding equation in static Universe). Last two equations admit simple exact determination of scalar field and potential energy density, while cosmological constant stands a "free parameter". Potential energy density represents a square function of scalar field with unique maximum (dynamically non-stable point). Any initial scalar field tends (co-exponentially) during time toward the same final scalar field, argument of the maximum of the potential energy density. It admits a possibility that final dynamically non-stable scalar field value turns out spontaneously in any other scalar field value when all begins again (like Sisyphus boulder motion).

It is well-known too that Linde chaotic inflation scenario [1]-[6] cannot be used for a consistent dynamical model of the dark energy or cosmological constant. (Simply speaking total energy density of the inflation field, oppositely to total energy density of the dark energy field [7], [8], is not constant.) But, in this work we shall suggest (in a formal analogy with Linde chaotic inflation scenario) simple dynamical model of the dark energy or cosmological constant. Concretely, we shall suggest an expression that can be formally considered as a Lagrangian dependent of Universe scale factor and spatially homogeneous (dark energy) scalar field (with strictly constant and positive total energy density that represents in fact cosmological constant). Then, corresponding Euler-Lagrange equation for Universe scale factor is formally exactly equivalent to the usual second Friedman equation for the flat empty space with (positive) cosmological constant (in this sense our model is full agreement with recent astronomical observations [7], [8]). Also there is corresponding Euler-Lagrange equation for scalar field that includes additional "friction" term and negative first partial derivative of unknown potential energy density (this equation is, in some way, similar to Klein-Gordon equation modified for cosmic expansion in Linde chaotic inflation scenario). Finally, total time derivative of the (constant) scalar field total energy density must be zero. It implies third dynamical equation which is formally exactly equivalent to usual Euler-Lagrange equation with positive partial derivative of unknown potential energy density (this equation is formally exactly equivalent to corresponding equation in static Universe). Last two equations admit simple exact determination of mentioned scalar field and potential energy density, while cosmological constant stands a free parameter. Potential energy density represents a square function of scalar field with unique maximum (dynamically non-stable point). Any initial scalar field tends (co-exponentially) during time toward the same final scalar field, argument of the maximum of the potential energy density. It represents very important fact. Namely, it admits a possibility (which will not be discussed here in details) that final dynamically non-stable scalar field turns out spontaneously and probabilistically in any other scalar field. Then dynamical evolution of the flat empty space with cosmological constant begins again. (It, in some way, corresponds to Linde eternal inflation concept, or, metaphorically, to Sisyphus boulder motion).

Introduce formally the following Lagrangian for Universe scale factor R

(1) $L = (dR/dt)^2/2 + (\rho/3)R^2/2 = (dR/dt)^2/2 + (\Lambda/3)R^2/2$

under formal condition

(2) $\rho = \Lambda = (d\varphi/dt)^2/2 + V = const > 0$

for a spatially homogeneous but time dependent scalar field $\varphi$ with unknown potential energy density V and total energy density $\rho$ identical to cosmological constant, where t represents time moment. (We use such units system in which speed of light c, reduced Planck constant $\hbar$ and $8\pi G$, where G represents Newtonian gravitational constant, equal 1.) In this way mentioned Lagrangian L (1) depends practically of two variables, explicitly of scale factor R and implicitly of (dark energy) scalar field. (Similar situation exists in Linde chaotic inflation scenario where Lagrangian represents product of the (inflation) scalar field Lagrangian density and volume of the sphere with radius R.) It can be observed that for non-unit $8\pi G$ expression (1) can be considered as the classical Lagrangian for unit mass system at surface of the sphere with radius R that gravitationally interacts with mentioned sphere homogeneously filled with mass with mass density $\rho$ (2).

Then corresponding Euler-Lagrange equation for R

(3) $d/dt\,(\partial L/\partial(dR/dt)) - \partial L/\partial R = 0$

or

(4) $d^2R/dt^2 - (\Lambda/3)R = 0$

corresponds to the second Friedman equation for flat empty space with cosmological constant $\Lambda$. This equation holds simple solution

(5) $R = R_0 \exp((\Lambda/3)^{1/2} t)$

where $R_0$ represents initial scale factor. Also, it simply implies the following first Friedman equation for flat empty space with cosmological constant $\Lambda$

(6) $(dR/dt)^2/R^2 = \Lambda/3 = \rho/3 = 1/3\,(1/2\,(d\varphi/dt)^2 + V)$

In this way both exact Friedman equations for flat, empty space with cosmological constant can be formally deduced by usual Lagrange formalism for scale factor. (It can be observed that first Friedman equation analogous to equation (6) exists in Linde chaotic inflation scenario. However in Linde scenario scalar field total energy density is only introduced in general relativistic first Friedman equation, but this equation is not deduced from any scalar field Lagrangian.) All this implies that our model is in full agreement with recent astronomical observations [7], [8] in sense that it does not ne3ed any revision of the general theory of relativity.

As it is not hard to see, total time derivative of (2) implies

(7) $d^2\varphi/dt^2 + \partial V/\partial\varphi = 0$

which has form equivalent to form of the usual Euler-Lagrange equation of the scalar field in static universe, i.e. for constant scale factor.

Further corresponding Euler-Lagrange equation for $\varphi$

(8) $d/dt\,(\partial L/\partial(d\varphi/dt)) - \partial L/\partial\varphi = 0$

for (1), (3) implies, as it is not hard to see,

(9) $d^2\varphi/dt^2 + 2(dR/dt)/R\,(d\varphi/dt) - \partial V/\partial\varphi = 0$

Obviously this equation holds, in respect to (7), not only additional "friction term" $2(dR/dt)/R\,(d\varphi/dt)$ , but also it holds opposite, negative sign before $\partial V/\partial\varphi$. (It can be observed that not completely analogous equation , Klein-Gordon equation modified for cosmic expansion, that represents Euler-Lagrange equation for (inflation) scalar field and mentioned Linde Lagrangian, exists in Linde chaotic inflation scenario. This equation has the following form $d^2\varphi/dt^2 + 3(dR/dt)/R\ (d\varphi/dt) + \partial V/\partial\varphi = 0$ so that it cannot have common non-trivial exact solutions with equation equivalent to (7) for any V. It represents main reason why Linde chaotic inflation scenario cannot be directly used for cosmological constant modeling.)

Now we can consider exists such V for which both (7) and (9) are exactly satisfied. (As it is well-known Linde chaotic inflation scenario is satisfied for a wide class of the scalar fields, i.e. scalar field potential energy densities. For this reason (9) can be simply transformed in

(10) $2d^2\varphi/dt^2 + 2(dR/dt)/R\ (d\varphi/dt) - (\ d^2\varphi/dt^2 + \partial V/\partial\varphi) = 0$

and then, according to (6) and (7), in

(11) $d^2\varphi/dt^2 + (\Lambda/3)^{1/2}\ (d\varphi/dt) = 0$

Simple solution of this equation holds the following general form

(12) $\varphi = A \exp(-(\Lambda/3)^{1/2}\ t) + B$

where A and B are some real constants. It implies

(13) $d\varphi/dt = -(\Lambda/3)^{1/2}\ (\varphi - B)$

(14) $d^2\varphi/dt^2 = (\Lambda/3)\ (\varphi - B)$

Introduction of (14) in (7) yields the following equation

(15) $(\Lambda/3)\ (\varphi - B) + \partial V/\partial\varphi = 0$

with unique simple solution

(16) $V = -\ (\Lambda/3)\ \varphi^2/2 + (\Lambda/3)\ B\ \varphi$

Obviously it represents a square function of the scalar field, atypical or forbidden in standard quantum scalar field theories. For positive $\Lambda$, potential energy density V (16) holds zero value for $\varphi=0$ and $\varphi=2B$ , while for $\varphi=B$ this V holds maximum or dynamically non-stable point $(\Lambda/3)\ B^2/2$ . According to (12), any finite scalar field initial value A+B (for non-zero A) tends (co-exponentially) toward the same scalar field final value B, i.e. toward dynamically non-stable point. It represents very important fact. Namely it admits a possibility (which will not be discussed here in details) that final dynamically non-stable scalar field value turns out spontaneously and probabilistically in any other scalar field value. Then described dynamical evolution of the flat empty space with cosmological constant begins again. (It, in some way, corresponds to Linde eternal inflation concept or, metaphorically, to Sisyphus boulder motion.)

Introduction of (13), (16) in (2) yields

(17) $\rho = \Lambda = (\Lambda/3)(\varphi - B)^2/2 - (\Lambda/3)\varphi^2/2 + (\Lambda/3) B \varphi = (\Lambda/3) B^2/2 = \mathrm{const}$ .

In this way important condition (2) is exactly satisfied. Moreover, (17) implies $B^2 = 6$ or $B = 6^{1/2}$ , but cosmological constant value stands non-determined "free parameter"..

Thus, it is definitely shown that suggested simple dynamical model of the dark energy or cosmological constant (deeply conceptually analogous, but not equivalent, to Linde chaotic inflation scenario) is theoretically consistent. It is based on formally introduced Lagrangian (1) dependent of two variables, scale factor and spatially homogeneous scalar field (with exactly constant and positive total energy density as cosmological constant). Corresponding Euler-Lagrange equation for scale factor (3) implies exactly Friedman equations for flat empty space with cosmological constant and in this sense our model is in full agreement with existing astronomical observations. Also, zero total time derivative of cosmological constant as the scalar field total energy density implies equation (7) which has form equivalent to form of the usual Euler-Lagrange equation of the scalar field in static universe, i.e. for constant scale factor.
Finally, corresponding Euler-Lagrange equation for scalar field (9) yields an especially modified equation. Last two equations have common non-trivial solution only for (co-exponentially) time dependent scalar field (12) and unique potential energy density as a square function of the scalar field (16). It means that any initial scalar field value tends (co-exponentially) during time toward the same final scalar field value that represents argument of the potential energy density unique maximum, i.e. dynamically non-stable point. It admits a possibility that final dynamically non-stable scalar field value turns out spontaneously in any other scalar field vačue when all begins again. (It, in some way, corresponds metaphorically to Sisyphus boulder motion). Cosmological constant value within our model stands theoretically non-determined "free parameter" but it, of course, does not represents any especial problem (since appearance of the theoretically "free parameters" which need experimental estimations represents almost universal characteristics in many field theories). A "small problem" appears in form of potential energy density (16) that includes term proportional to scalar field (absent in typical quantum field theories). But without mentioned term suggested model is either inconsistent or trivial (with zero cosmological constant). It implies that suggested model is an effective, hybrid model that needs further generalizations.